\documentclass[aps, pre, floatfix,
showpacs,
twocolumn
]{revtex4-1}

\usepackage{graphicx}
\usepackage{bm}
\usepackage{subfigure}
\usepackage{hyperref}
\usepackage{amsmath}
\usepackage{color}

\begin{document}

\title{Algorithms to generate saturated random sequential adsorption packings built of rounded polygons}

\author{Micha\l{} Cie\'sla}
    \email{michal.ciesla@uj.edu.pl}
\author{Piotr Kubala}
    \email{piotr.kubala@doctoral.uj.edu.pl}
\author{Konrad Kozubek}
    \email{konrad.p.kozubek@gmail.com}

\affiliation{Institute of Theoretical Physics, Department of Statistical Physics, Jagiellonian University, \L{}ojasiewicza 11, 30-348 Krak\'ow, Poland}


\date{\today}

\begin{abstract}
We present the algorithm for generating strictly saturated random sequential adsorption packings built of rounded polygons. It can be used to study various properties of such packings built of a wide variety of different shapes and in modelling monolayers obtained during the irreversible adsorption processes of complex molecules. Here, we apply the algorithm to study the densities of packings built of rounded regular polygons. Contrary to packings built of regular polygons, where packing fraction grows with an increasing number of polygon sides, the packing fraction reaches its maximum for packings built of rounded regular triangles. With a growing number of polygon sides and increasing rounding radius, the packing fractions tend to the limit given by a packing built of disks. However, they are still slightly denser, even for the rounded 25-gon, which is the highest-sided regular polygon studied here.
\end{abstract}


%
\maketitle
\section{Introduction}
Random packings are a natural model of random media, whose properties are an active area of study due to their fundamental and utilitarian applications \cite{Lee1991, Truskett1998,Torquato2009, Torquato2010, Agarwal2011}. There are a lot of different types of random packings. The most popular ones are random close packings (RCP), where the process used to form the packing is continued until packing density stops growing. There, the neighbouring particles are in contact. RCP is mainly used in the modelling of granular matter \cite{Jaeger1992, Meng2016, Liu2018}. However, despite the prominence of this type of packing, it appears that the very notion of RCP is ill-defined because one cannot simultaneously maximise packing density and packing disorder \cite{torquato2000}. Moreover, the packing density is sensitive to the details of the numerical or even experimental protocol used \cite{Asencio2017}. Another kind of packing, which has well defined mean packing density, is obtained using random sequential adsorption (RSA) protocol \cite{Evans1993}. The packing is created using the following iterative scheme:
\begin{itemize}
    \item a virtual particle of random position and orientation within a packing is selected;
    \item if the virtual particle does not intersect with any of the particles in the packing, it is added to the packing, and its position and orientation remain fixed;
    \item otherwise, the virtual particle is removed and abandoned.
\end{itemize}
This procedure should be continued until the packing is saturated, which means there is no place large enough to add another object.

Although RSA history began early in the past century \cite{Flory1939}, it was popularised by Feder, who noticed that such packings resemble monolayers built during irreversible adsorption processes \cite{Feder1980}. In these processes, the molecules from a bulk phase attach at random places to a surface (or an interface) and form a layer. The structure of such a layer is determined mainly by a geometrical cross-section of molecule shape and two-dimensional surface. Therefore, adsorption layers are well modelled by two-dimensional loose random packings produced by the RSA algorithm. Comparing experimentally measured characteristics to ones obtained from such modelling, allows one to check if assumed interactions between the particles adsorbed and the surface influencing molecule orientation in the layer, are correct \cite{Hemmersam2008, Min2017, Manzi2019, Kosior2020}. In such studies, a cross-section of a complex molecule shape with a surface is typically approximated by simple geometrical objects like disks, two-dimensional spherocylinders or ellipses \cite{Adamczyk2006, Matijevic2004}. For more complex molecules, models may use several disks \cite{Adamczyk2010, Ciesla2013, Kujda2015}. However, this approach follows cross-sections that are not smooth and concave, introducing unwanted steric effects. This study presents an effective method of generating RSA packings built of arbitrary polygons with rounded corners. Such an object can model virtually any two-dimensional cross-section used in the numerical modelling of adsorption processes.

It is worth mentioning that besides these practical applications, RSA plays the role of one of the simplest, yet not trivial protocols of forming disordered packings which account for excluded volume effects \cite{Onsager1932, Baule2017, Baule2019, Yuan2019}. Therefore, our algorithm can also be used in this area, to study properties of a wide range of different shapes. On the other hand, RSA cannot be used to study other phenomena induced by the excluded volume, e.g. phase transitions \cite{Onsager1949,Frenkel1987}, since the system it is essentially non-equilibrium \cite{Torquato2018}.

The main problem that has to be addressed by implementing the RSA algorithm is its performance near saturation. While in adsorption experiments, saturated monolayers are typically obtained in minutes, the standard approach to the RSA algorithm becomes highly ineffective when the probability of finding a large enough place for another particle is tiny. Therefore, the expected number of RSA iterations needed to draw this place is very large, which leads to the very long simulation times. Additionally, even when saturation is reached and there is no possibility of adding subsequent object to the packing, the algorithm still tries to add the next randomly placed shape, because it cannot detect that the packing is already saturated. These problems have been solved only for several, specific shapes, for example disks \cite{Wang1994, Ebeida2012, Zhang2013}, and recently for ellipses and spherocylinders \cite{Haiduk2018}, polygons \cite{Kasperek2018, Zhang2018, Ciesla2019} and figures built of several disks \cite{ciesla2020}. This paper extends this list by showing how to generate saturated RSA packings built of polygons with rounded corners. The algorithm proposed is tested by studying the properties of RSA packings' built of rounded regular polygons.
\section{Saturated packing generation}
\begin{figure}[htb]
  \centering{
    \includegraphics[width=0.6\columnwidth]{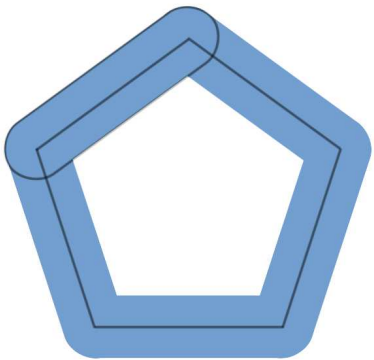}
  }
  \caption{Illustrative rounded pentagon, where segments (black lines) are replaced with (blue) spherocylinders.}
  \label{fig:roundedpentagon}
\end{figure}

\begin{figure}
    \centering{
    \includegraphics[width=0.6\linewidth]{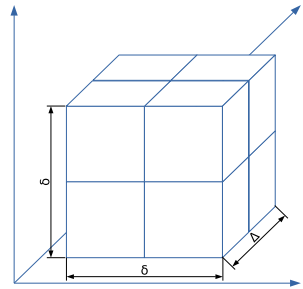}
    }
    \caption{Division of the voxel of spatial size $\delta$ and angular size $\Delta$ to 8 smaller voxels of spatial size $\delta/2$ and angular size $\Delta/2$.}
    \label{fig:voxeldivision}
\end{figure}

\begin{figure}
    \centering{
    \includegraphics[width=0.9\linewidth]{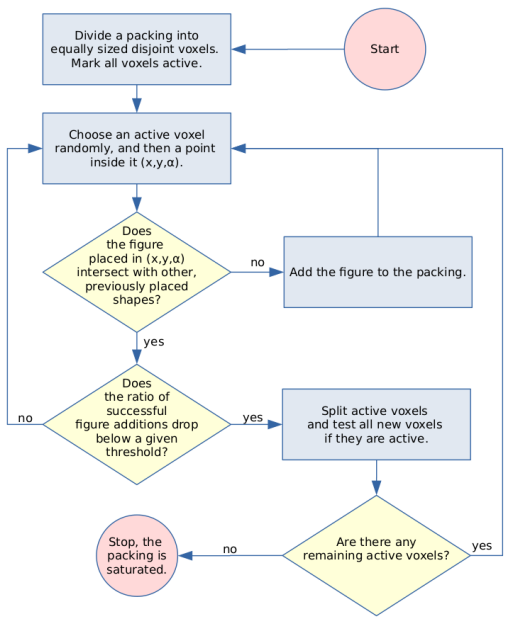}
    }
    \caption{Block diagram of the algorithm that generates strictly saturated RSA packing.}
    \label{fig:rsa-scheme}
\end{figure}

First, let us note that the rounded polygon, instead of segments, is built of spherocylinders -- see Fig.~\ref{fig:roundedpentagon}.
To generate a strictly saturated packing built of such shapes, we follow the idea of tracking the space, where subsequent particles can be placed \cite{Wang1994,  Ebeida2012, Zhang2013, Brosilow1991, Torquato2006}. At the beginning of the packing generation, the whole packing is divided into disjoint voxels. Each voxel is defined using five numbers. Three of them are its coordinates $(x_v, y_v, \alpha_v)$, and extra two $(\delta, \Delta)$ denote its size. Thus, a voxel is a set of points $(x, y, \alpha)$ such that $x \in [x_v - \delta/2,\, x_v + \delta/2)$, $y \in [y_v - \delta/2,\, y_v + \delta/2)$ and $\alpha \in [\alpha_v - \Delta/2,\, \alpha_v + \Delta/2)$. Each point inside a voxel corresponds to the centre and the orientation of the trial rounded polygon. A crucial part of the algorithm is to determine if the given voxel contains at least one point that corresponds to a polygon, which does not intersect with any polygon already added to the packing. If there is no such a point, the voxel becomes inactive. Note that inactive voxel cannot become active again, because shapes added to the packing neither vanish, nor  change their positions, thus there is no possibility that in such a voxel a point corresponding to non intersecting trial object will appear at further stages of packing generation. Inactive voxels are not used in the sampling of trial objects, which speeds up packing generation, therefore they can be removed to lower the usage of computer memory. If all voxels become inactive or removed, the packing is saturated -- there is no possibility to add a new object to the packing. Note that active voxels estimate regions, where new figures can be added. The smaller the voxel size (parameters $\delta$ and $\Delta$), the better the approximation. However, voxels cannot be arbitrarily small from the beginning of the packing generation due to computer memory limit. Therefore, typically, calculations are started with relatively large voxels. When the ratio of unsuccessful trials of adding a new figure to the packing exceeds some threshold value, all active voxels are divided -- each voxel into eight smaller ones -- see Fig.~\ref{fig:voxeldivision}, and then all these new voxels are tested if they will remain active. 
The threshold value mentioned may be smaller when there is a lot of memory available for calculations, but should be large when we want to avoid too frequent voxel divisions to keep their number limited. The block diagram of the complete algorithm that generates a saturated packing is shown in Fig.~\ref{fig:rsa-scheme}.
We have developed two different ways to test whether the voxel should be marked inactive. The first one is based on algebraic worst-case estimation, while the second one uses the geometric notion of excluded zones. Both are described in the next section.

\subsection{Voxel elimination}

\subsubsection{Algebraic approach}

\begin{figure}[htb]
  \centering{
    \includegraphics[width=0.7\columnwidth]{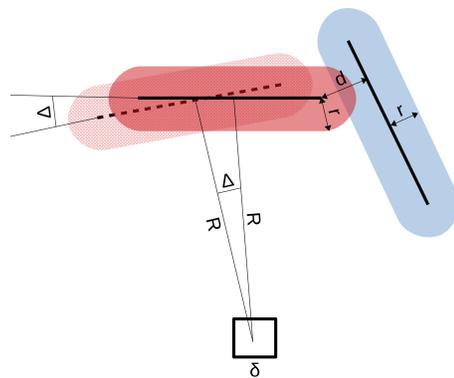}
  }
  \caption{The illustrative red spherocylinder from a trial particle placed in the voxel's centre and the blue spherocylinder from a particle in the packing. The distance between these spherocylinders varies depending on a particular point in the voxel where the trial particle's centre is placed and the particle's orientation.}
  \label{fig:voxelelimination}
\end{figure}

First, we notice, that the spherocylinder is fully defined using a given segment and the maximal distance between this segment and the set of points that build the spherocylinder. Thus, for a voxel to be marked inactive, the distance between the segments giving two spherocylinders: one belonging to a trial particle and the second one to a polygon already placed inside the packing, has to be smaller than the spherocylinder width $2r$, regardless of the centre and the orientation of the trial particle inside the voxel. Therefore, first, we use Zhang criterion \cite{Zhang2018} to determine if it is possible to avoid these segments' intersection. If so, we have to estimate the minimal possible distance between them. To do this, assume, that for an object placed inside a voxel centre $(x_v, y_v, \alpha_v)$ the distance between the segments is $d$. Due to the rotation inside the voxel, the red spherocylinder will not move further than $2 R \sin (\Delta/2)$, where $R$ is the radius of the polygon's circumscribed circle, and due to the translation -- than $(\sqrt{2}/2)\delta$ -- see Fig.~\ref{fig:voxelelimination}. 
Thus, if
\begin{equation}
    d + 2 R \sin \left( \frac{\Delta}{2} \right) + \frac{\sqrt{2}}{2}\delta < 2r,
    \label{eq:voxelelimination}
\end{equation}
the spherocylinders have to intersect, and therefore the voxel can be marked as inactive. Otherwise, the voxel remains active, even if it cannot host any non-intersecting polygon centre, but the above, rough estimation is satisfied. However, note that when voxel size tends to $0$, condition (\ref{eq:voxelelimination}) approaches the spherocylinders intersection condition. Therefore, for decreasing size of the voxels, the accuracy of sampling non-intersecting shapes grows up.

\begin{figure}[htb]
  \centering{
    \includegraphics[width=0.7\columnwidth]{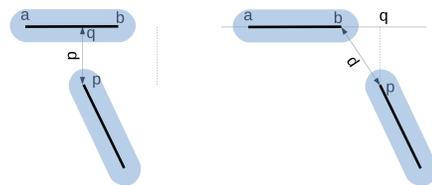}
  }
  \caption{Determination of distance $d$ between point $p$ and segment $ab$. Point $q$ is the orthogonal projection of the point $p$ onto the line containing $a$ and $b$. If $q$ is between $a$ and $b$ (left panel), the distance between the point $p$ and the segment equals $|pq|$. Otherwise (right panel), it equals $\min \{ |pa|, |pb| \}$.}
  \label{fig:pointsegment}
\end{figure}

The last remaining point is to determine the distance between two, non-intersecting segments. Here, we calculate four distances from the first segment's ends to the second and vice-versa (see Fig.~\ref{fig:pointsegment}).
The smallest of these values is the distance between two segments. Two rounded polygons do not intersect if the distances between all their segments are larger than $2r$.

\subsubsection{Geometric approach}

\begin{figure*}
	\centering{
	\subfigure[]{\includegraphics[width=0.63\linewidth]{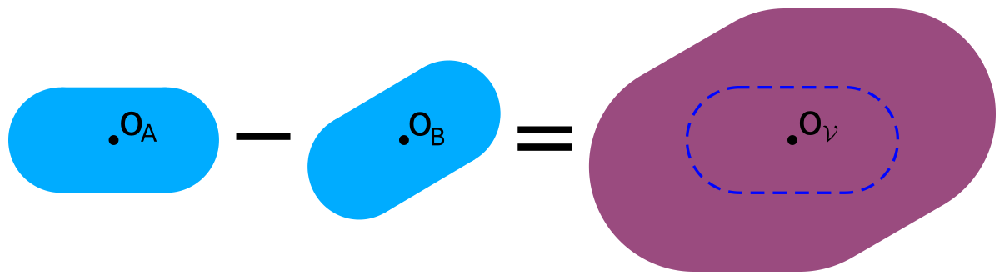}}
	\hspace{30pt}
	\subfigure[]{\includegraphics[width=0.27\linewidth]{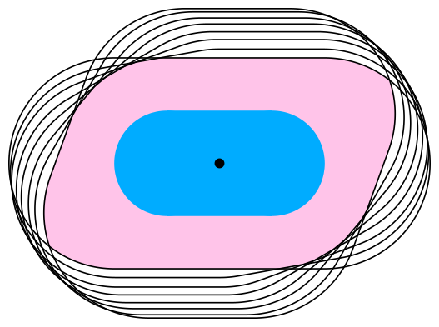}}
	\subfigure[]{\includegraphics[width=0.63\linewidth]{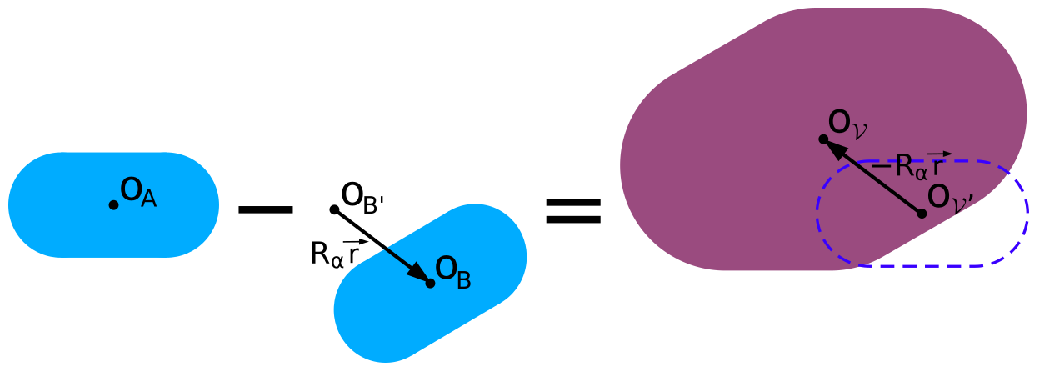}}
	\hspace{30pt}
	\subfigure[]{\includegraphics[width=0.27\linewidth]{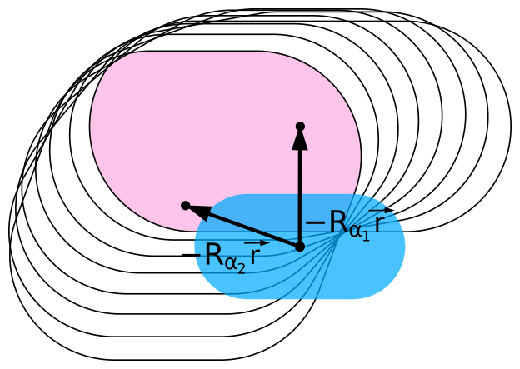}}
	\subfigure[]{\includegraphics[width=0.27\linewidth]{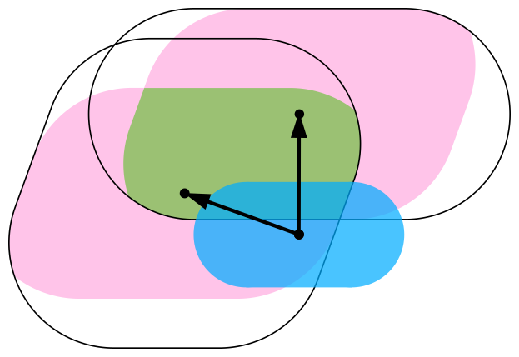}}
	\hspace{30pt}
	\subfigure[]{\includegraphics[width=0.63\linewidth]{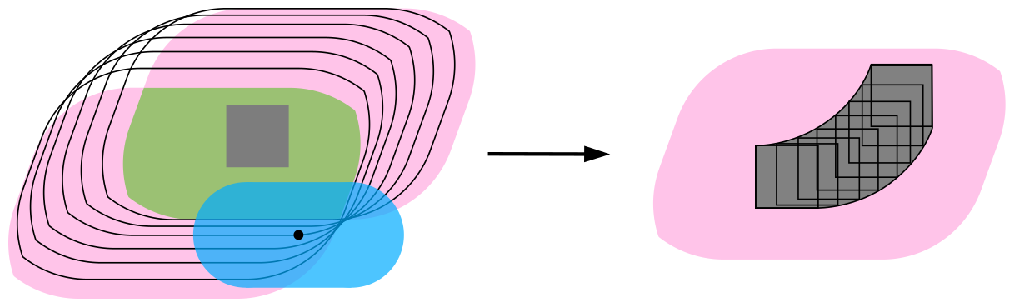}}
	}
	\caption{(a) Illustration of the exclusion zone $\mathcal{V}(\alpha)$ for two spherocylinders $A$, $B(\alpha)$. The dashed line represents $A$. $O_X$ denotes the origin of the respective geometric object $X$ -- in particular, $O_\mathcal{V}$ is the origin of the exclusion zone $\mathcal{V}(\alpha)$. (b) A few exclusion zones for a spherocylinder in the range $[\alpha_1, \alpha_2)$ (solid lines) and the intersection of them (coloured area). (c) Analogous exclusion zone construction for a shifted spherocylinder $B'$. Primed symbols correspond to shifted shapes. (d) A few of exclusion zones for $B'$. The arrows point at centres of exclusion zones for endpoints of $[\alpha_1, \alpha_2)$. Notice how all exclusion zones are shifted with respect to $A$ by an angle-dependent vector $-R_\alpha \vec{r}$. (e) Illustration of approximating $\mathcal{V}'(\alpha)$ by $\mathcal{V}_\cap$. Pink (lighter) areas represent $\mathcal{V}_\cap$ for endpoints of $[\alpha_1, \alpha_2)$, solid lines are full $\mathcal{V}'(\alpha)$, while green (darker) area represents the intersection of all $\mathcal{V}_\cap$. The last one is the final exclusion zone. (f) An example voxel to be marked inactive because it lies inside each $\mathcal{V}_\cap$. It is equivalent to examining a ``smeared'' voxel $V_\cup$ and a single exclusion zone $\mathcal{V}_\cap$. Coloured areas coincide with those from (e), while solid lines are $\mathcal{V}_\cap$ approximating a few intermediate $\mathcal{V}'(\alpha)$.}
    \label{fig:exclusion_zones}
\end{figure*}

We start by introducing the exclusion zone $\mathcal{V}$ of a particle $A$ with an origin $O_A$ for a particle $B$ with an origin $O_B$, which is a subset of space, for which $B$ overlaps $A$ if and only if $O_B$ lies inside $\mathcal{V}$. Mathematically, it is defined as the Minkowski difference of $A$ and $B$, namely
\begin{equation}
    \mathcal{V} = A - B \equiv \{\vec{a}-\vec{b}:\: \vec{a} \in A,\, \vec{b} \in B\},
\end{equation}
where $\vec{a}$ and $\vec{b}$ represent vectors pointing from the particle's origin to one of its points, for $A$ and $B$ respectively.

To begin with, let $A$ and $B$ be an arbitrary pair of spherocylinders, whose origins correspond to their geometrical centres. We assume that the variable orientation of a trial spherocylinder relative to some axis $\alpha$ is $B(\alpha)$. We denote the exclusion zone of $A$ for $B(\alpha)$ as $\mathcal{V}(\alpha)$ (see Fig.~\ref{fig:exclusion_zones}a). Note that $\mathcal{V}(\alpha)$ is convex because it is the Minkowski difference of two convex sets. A voxel $V$ can be marked inactive if its spatial part is contained in the intersection $\mathcal{V}_\cap(\alpha_1,\alpha_2)$ of all exclusion zones for the angles from the range $[\alpha_1, \alpha_2) = [\alpha_v - \Delta/2, \alpha_v + \Delta/2)$:
\begin{equation} \label{eq:max_exzone}
    \mathcal{V}_\cap(\alpha_1,\alpha_2) = \bigcap_{\alpha_1 \leq \alpha < \alpha_2} \mathcal{V}(\alpha).
\end{equation}
$\mathcal{V}_\cap$ can be determined for spherocylinders using elementary geometrical consideration, which was described in \cite{Haiduk2018}. It is illustrated in  Fig.~\ref{fig:exclusion_zones}b. There also exist similar constructions for ellipses and rectangles \cite{Haiduk2018, Kasperek2018}.

However, the origins of spherocylinders building rounded polygons coincide in a common point different from their geometric centres. Let $A$ be a fixed spherocylinder and $B'(\alpha) = B(\alpha) + R_\alpha \vec{r}$ a spherocylinder from a trial particle, displaced from the common origin by $R_\alpha \vec{r}$, where $\vec{r}$ denotes the displacement for $\alpha=0$ and $R_\alpha$ is the rotation operator. The new exclusion zone reads as (Fig.~\ref{fig:exclusion_zones}c)
\begin{align}
    \mathcal{V}'(\alpha) &= A - B'(\alpha) \notag \\ 
    &= A - B(\alpha) - R_\alpha \vec{r} \notag \\
    &= \mathcal{V}(\alpha) - R_\alpha \vec{r},
\end{align}
so a voxel can be marked inactive if
\begin{equation}
    V \subset \bigcap_{\alpha_1 \leq \alpha < \alpha_2} \mathcal{V}(\alpha) - R_\alpha \vec{r}.
\end{equation}
The shape of the intersection of exclusion zones is more complex in this case and depends more subtly on $\alpha_1$ and $\alpha_2$ (see Fig.~\ref{fig:exclusion_zones}d). We can mitigate the issue by making an appropriate approximation -- we approximate all $\mathcal{V}(\alpha)$ by the intersection $\mathcal{V}_\cap(\alpha_1,\alpha_2)$:
\begin{equation}
    V \subset \bigcap_{\alpha_1 \leq \alpha < \alpha_2} \mathcal{V}_\cap(\alpha_1, \alpha_2) - R_\alpha \vec{r}.
\end{equation}
This step is illustrated in Fig.~\ref{fig:exclusion_zones}e. From the construction, this approximation does not introduce false-positive voxel rejections. Next, we notice that this is equivalent to checking if
\begin{equation}
    V_\cup(\alpha_1, \alpha_2) \subset \mathcal{V}_\cap(\alpha_1, \alpha_2),    
\end{equation}
where
\begin{equation}
    V_\cup(\alpha_1, \alpha_2) = \bigcup_{\alpha_1 \leq \alpha < \alpha_2} V + R_\alpha \vec{r}.
\end{equation}
It means that we check if $\mathcal{V}_\cap(\alpha_1, \alpha_2)$ contains a voxel ``smeared'' on an arc (see Fig.~\ref{fig:exclusion_zones}f). Because in the case of spherocylinders, both $\mathcal{V}_\cap$ and $V_\cup$ boundaries consist solely of arcs and segments, the condition for voxel rejection can be evaluated without further approximations. However, one can also simplify it using a rectangular bounding of $V_\cup$. As both the bounding and $\mathcal{V}_\cap$ are convex, it is sufficient to check whether the vertices of the bounding lie inside $\mathcal{V}_\cap$. In either case, the procedure converges to the intersection criterion for a voxel size tending to 0.
\section{Numerical simulations}
To test the above-described algorithms, we studied saturated packings built of several rounded regular polygon types. These polygons were built of $n = 3, 4, 5, 6, 7, 8, 9, 10, 15, 25$ segments. The circumscribed circle radius was $1+r$ where $r$ varied from $0.2$ to $2.0$ and was equal to a half-width of spherocylinders replacing polygon segments. These shapes were placed according to the RSA protocol on a square, whose surface area was $10^6$ times larger than the surface area of a single rounded polygon. We used periodic boundary conditions to decrease finite-size effects \cite{Ciesla2018}. For each shape studied, up to a $100$ independent random packings were generated to estimate the mean packing fraction. It allowed us to keep statistical error slightly below $2.0 \times 10^{-5}$, which was enough to compare packing fractions for different shapes. We used  the algebraic approach to eliminate inactive voxels, as in this case, it was significantly faster than the second method based on exclusion zones. The comparison of the speed of the two approaches is presented in the latter part of the manuscript. 

Here, it is worth to note that the speed of packing generation depends on several parameters. The crucial one is the number of consecutive iterations without adding a particle which triggers voxel division. The optimal value of this parameter depends on a packing size, the shape of deposited objects and the maximal number of voxels (limited by available RAM), thus it is hard to be estimated {\em a priori}. Too low values of this parameter cause frequent divisions, resulting in problems with voxel storage and extended time needed for their analysis, while too large values cause too many unsuccessful tries of adding a particle to the packing. In this study, the first division occurred typically after $10^4 - 10^5$ of such iterations and then, to maintain a similar average number of tries for each voxel, we adjusted the parameter to be proportional to the number of active voxels. However, for different shapes or packing sizes, not studied here, we found the optimal value of this parameter ranges from $10^2$ up to $10^6$.

Another parameter is the initial size of a voxel. Too small voxels overload computer memory, while too large ones are harder to deactivate and remove. Here we started the simulation according to the standard RSA procedure. Then, after $10^4 - 10^5$ unsuccessful consecutive tries we initialised voxel structure. The diagonal of each voxel did not exceed the diameter of the circle inscribed in the studied shape. It is a maximal spatial size that guarantees voxel deactivation after placing a shape in it. The angular size was equal $\Delta = 0.25$. Due to computer representation of floating point numbers, to avoid problems with numerical precision, it is worth to use voxels of spatial and angular sizes equal to a power of $2$. Additionally, for symmetrical particles like rounded polygons, to reduce the total number of voxels, the range of possible trial angles can be limited from $[0, 2\pi)$ to $[0, 2\pi / n)$ where $n$ is the number of polygon sides.
\section{Results and discussion}
\begin{figure}[htb]
  \centering{
  \subfigure[]{\includegraphics[width=0.3\columnwidth]{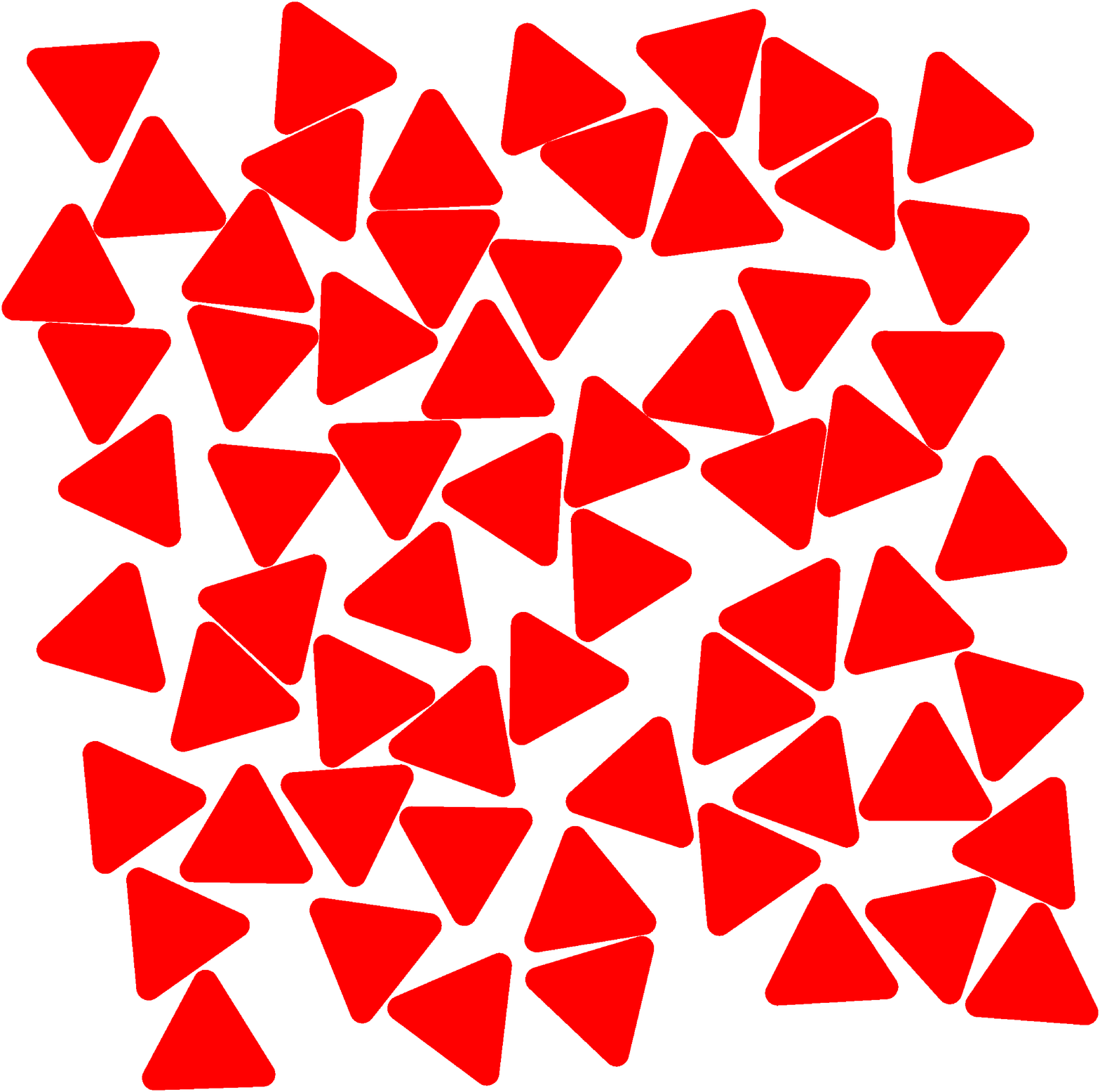}}
  \subfigure[]{\includegraphics[width=0.3\columnwidth]{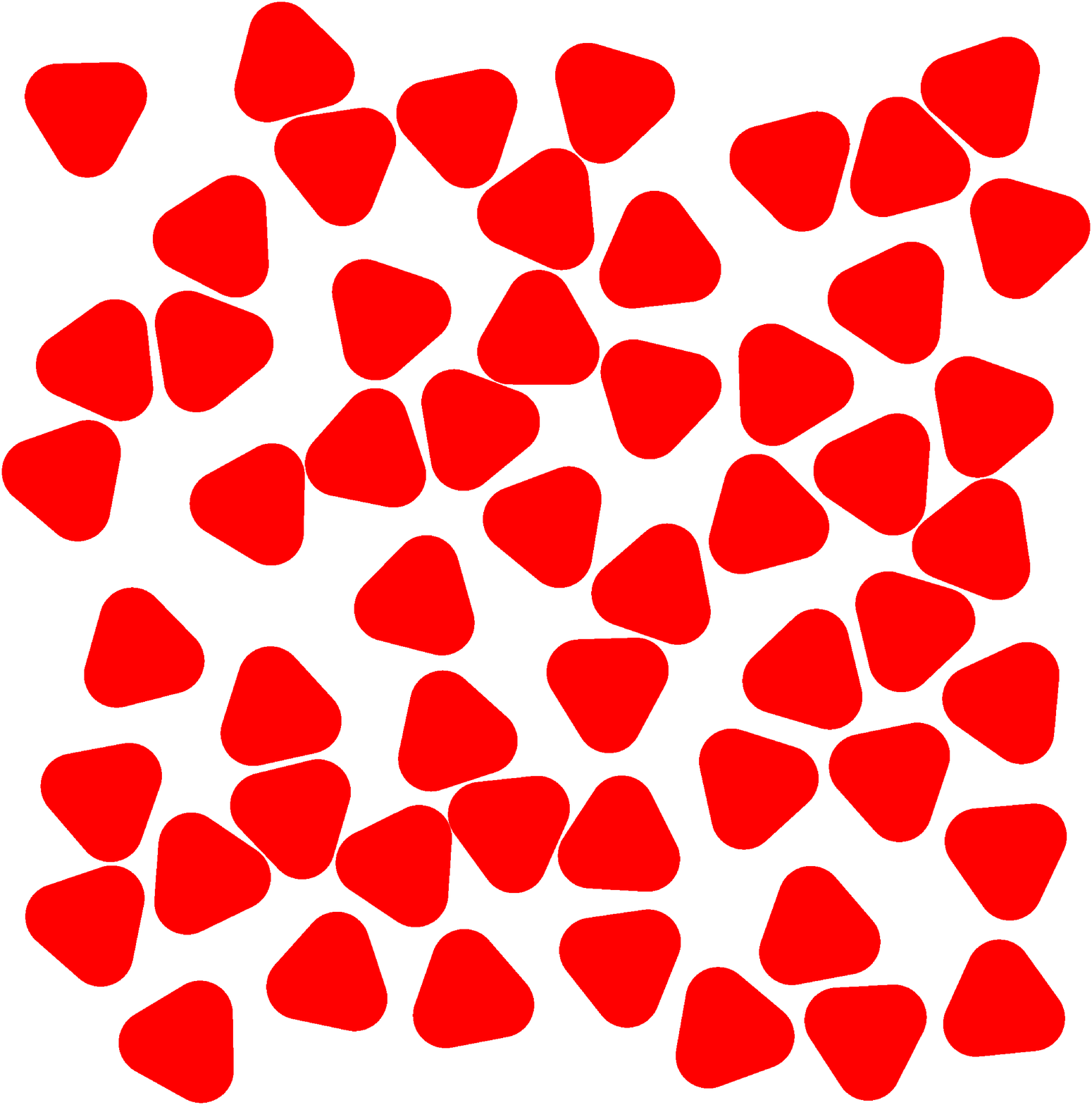}}
  \subfigure[]{\includegraphics[width=0.3\columnwidth]{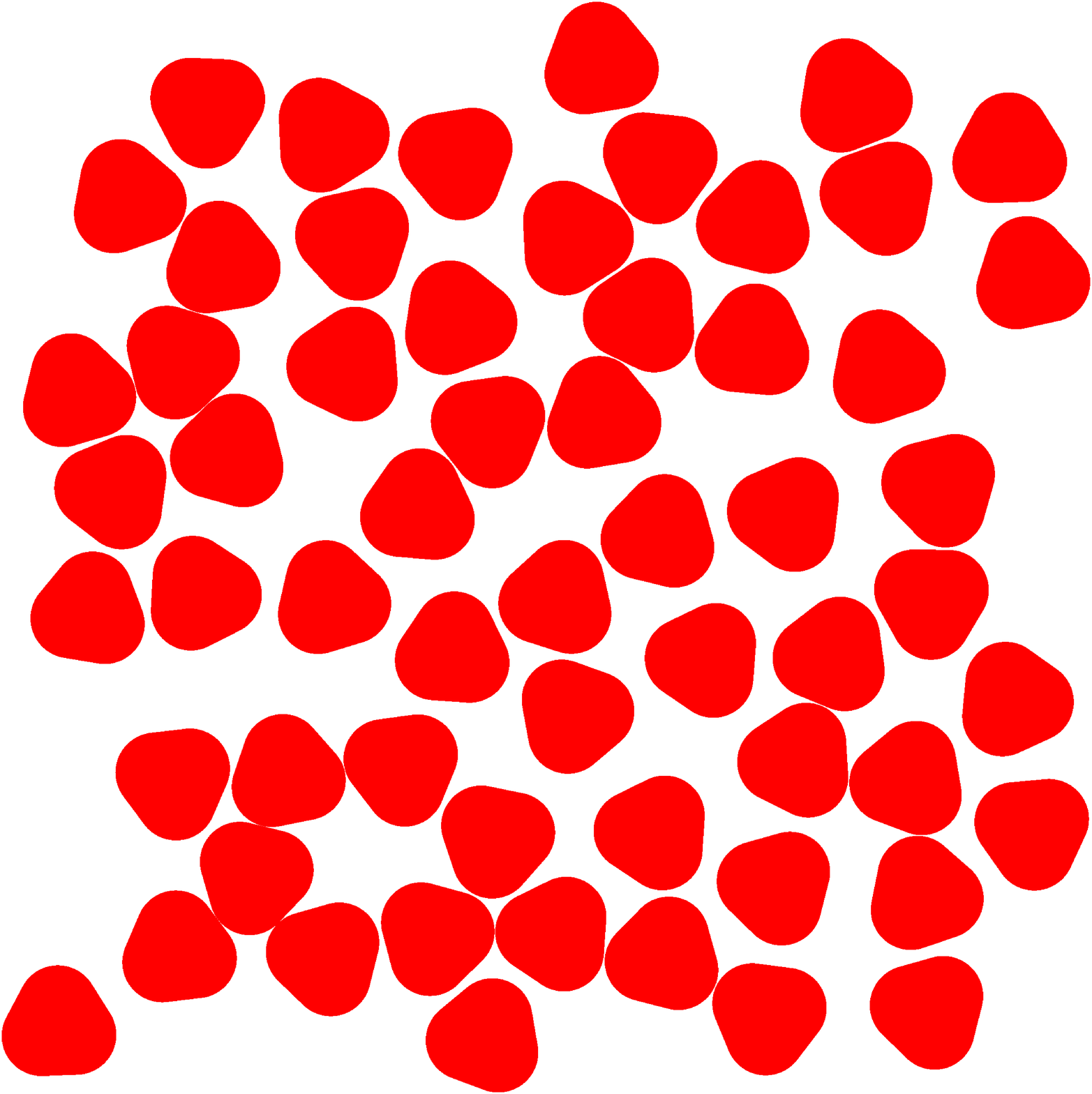}}
  }
  \caption{Illustrative, small saturated random packings built of rounded equilateral triangles for $r = 0.2, 1.0\text{ and }2.0$ for panels (a), (b) and (c), respectively. Periodic boundary conditions are used.}
  \label{fig:triangles}
\end{figure}

\begin{figure}[htb]
  \centering{
  \subfigure[]{\includegraphics[width=0.3\columnwidth]{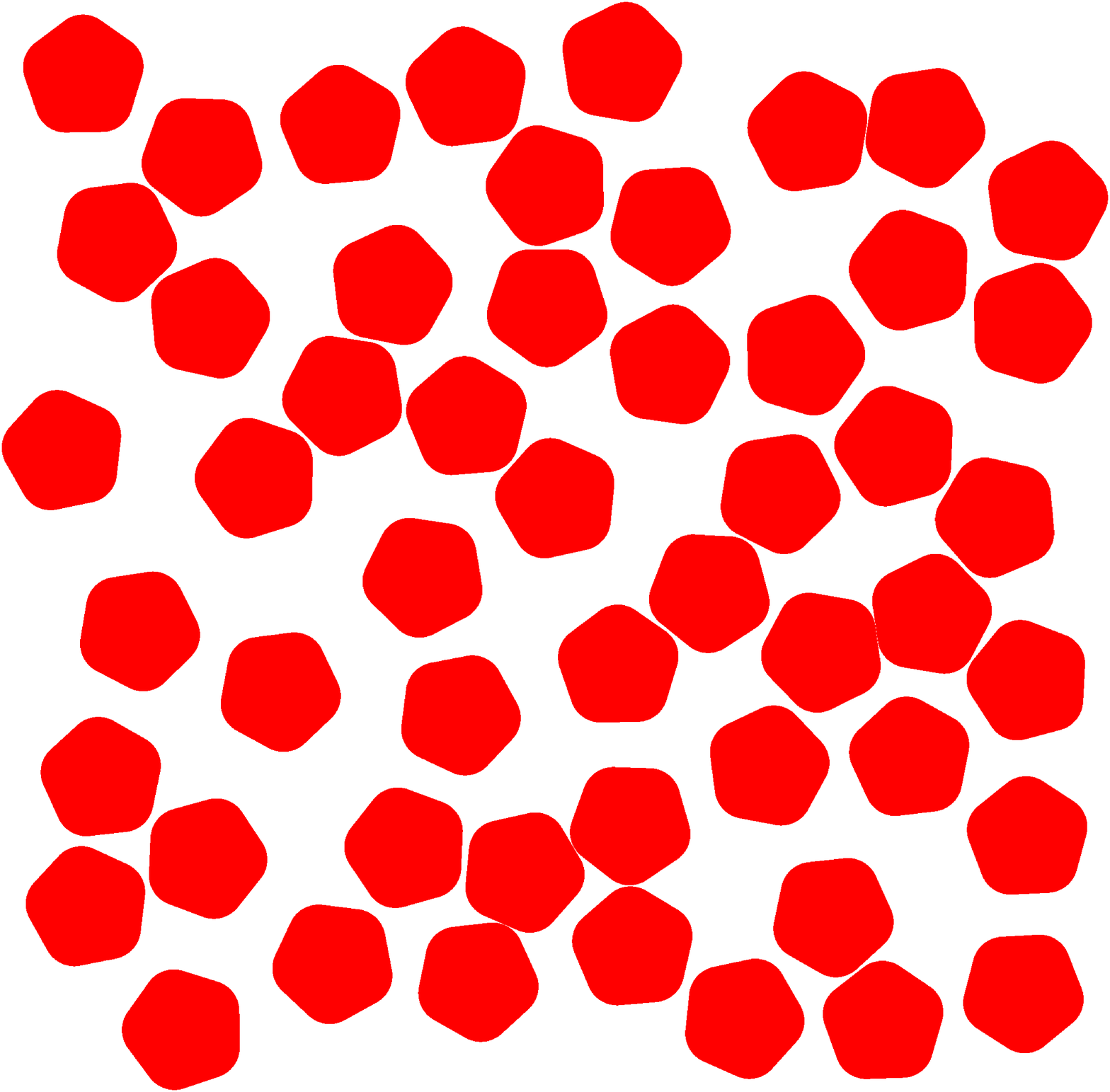}}
  \subfigure[]{\includegraphics[width=0.3\columnwidth]{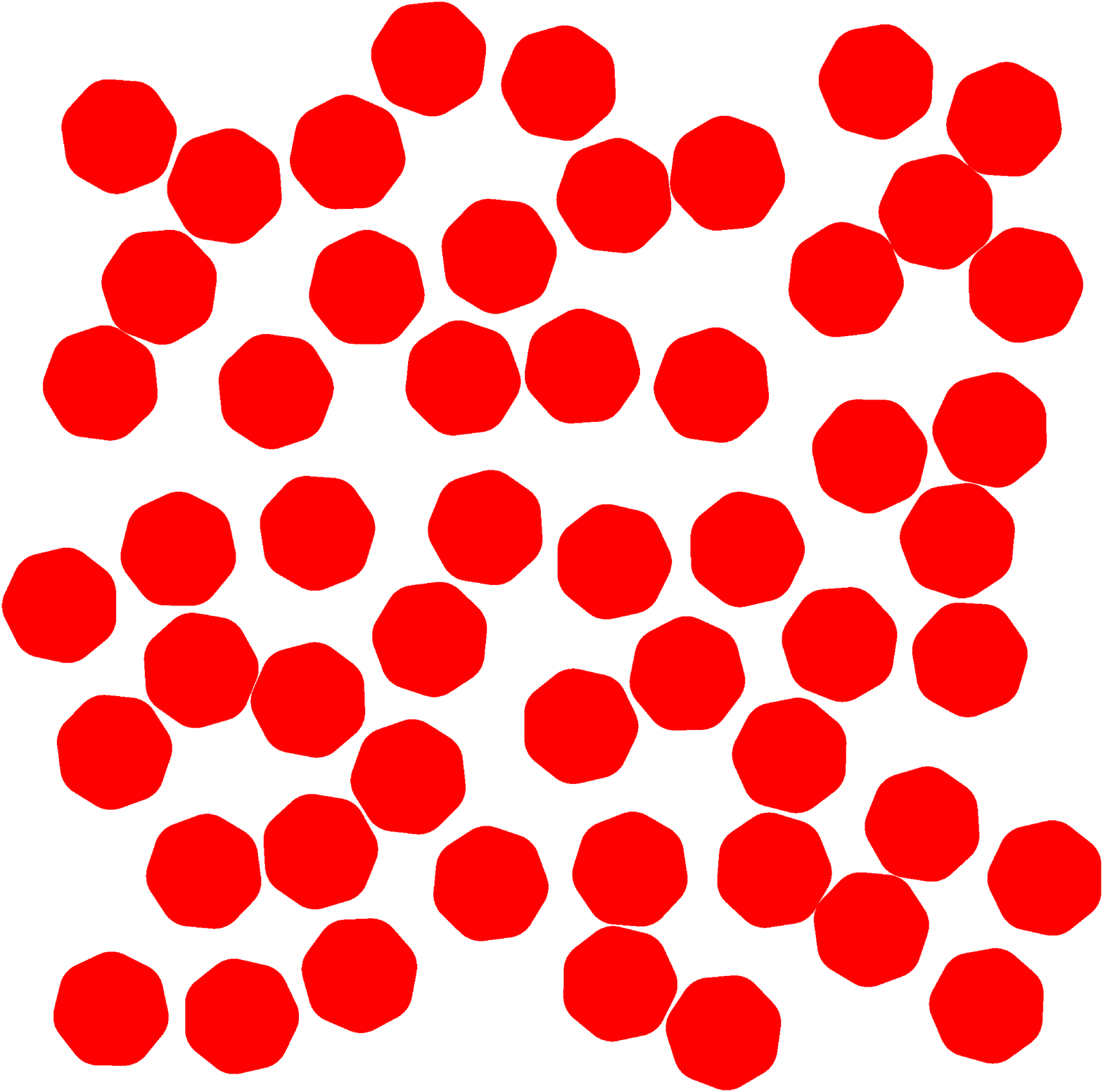}}
  \subfigure[]{\includegraphics[width=0.3\columnwidth]{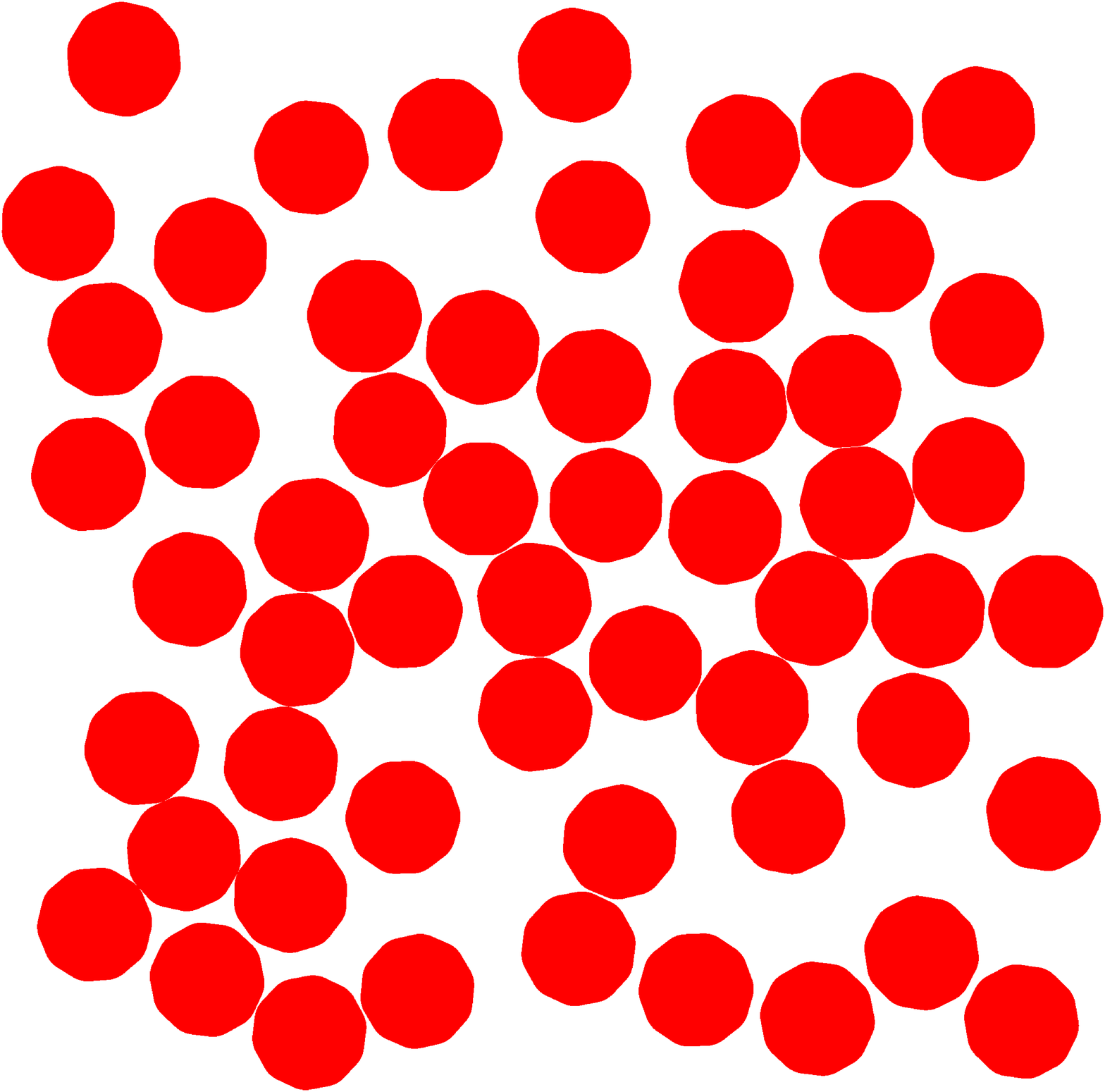}}
  }
  \caption{Illustrative, small saturated random packings built of (a) rounded pentagons, (b) heptagons and (c) decagons for $r = 1.0$. Periodic boundary conditions are used.}
  \label{fig:5710gons}
\end{figure}

Illustrative, saturated packings are shown in Figs.~\ref{fig:triangles} and \ref{fig:5710gons}. 
We firstly compared our results for $r=0$ with the ones reported in the study of regular polygons \cite{Zhang2018}, and they agree within statistical error. For example, for the regular pentagon, the most accurate estimation of the packing fraction so far is $\theta=0.541344 \pm 0.000072$, while our simulations for a similar packing size follow to $\theta=0.541190 \pm 0.000080$.

\begin{figure}[htb]
  \centering{
  \subfigure[]{\includegraphics[width=0.49\columnwidth]{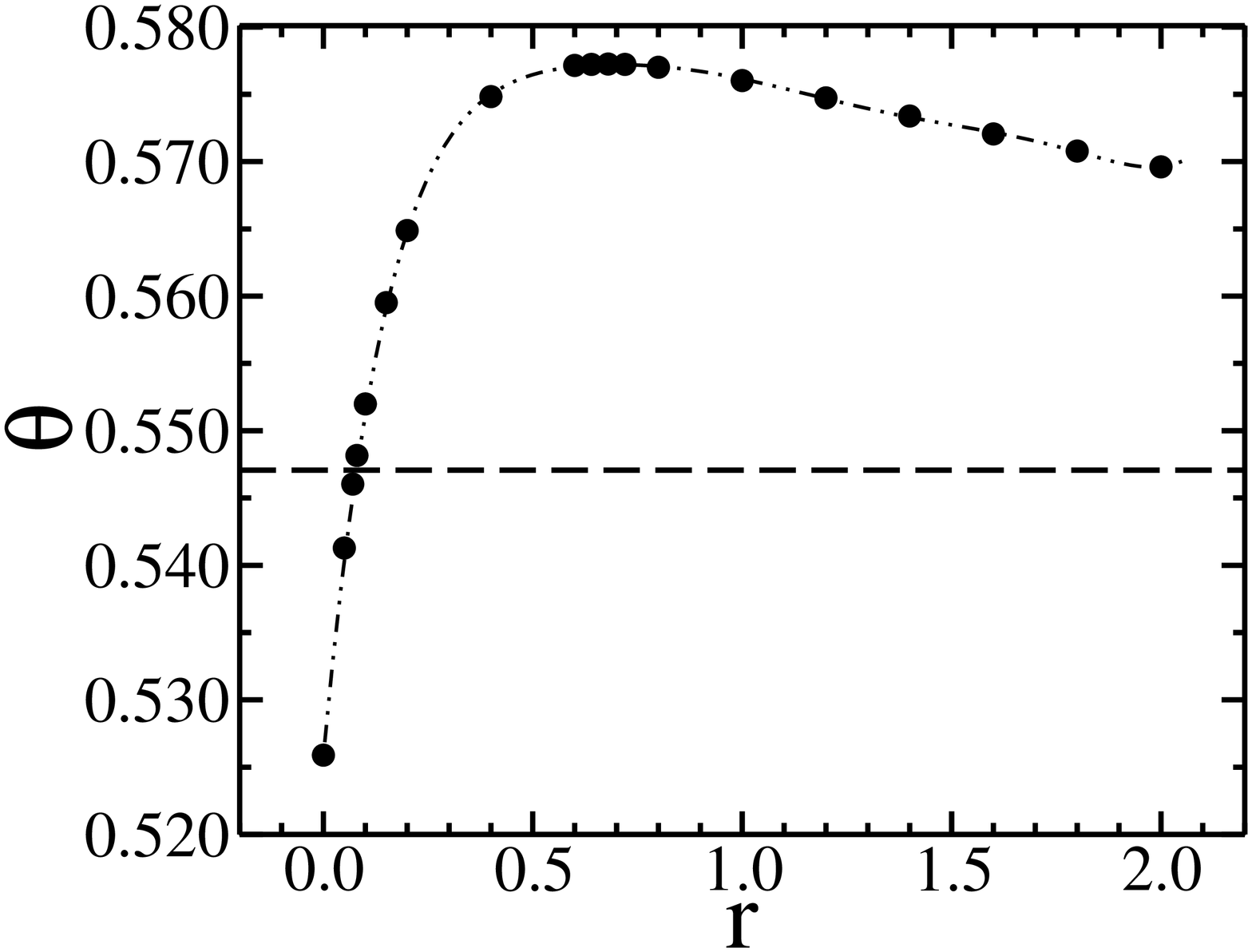}}
  \subfigure[]{\includegraphics[width=0.49\columnwidth]{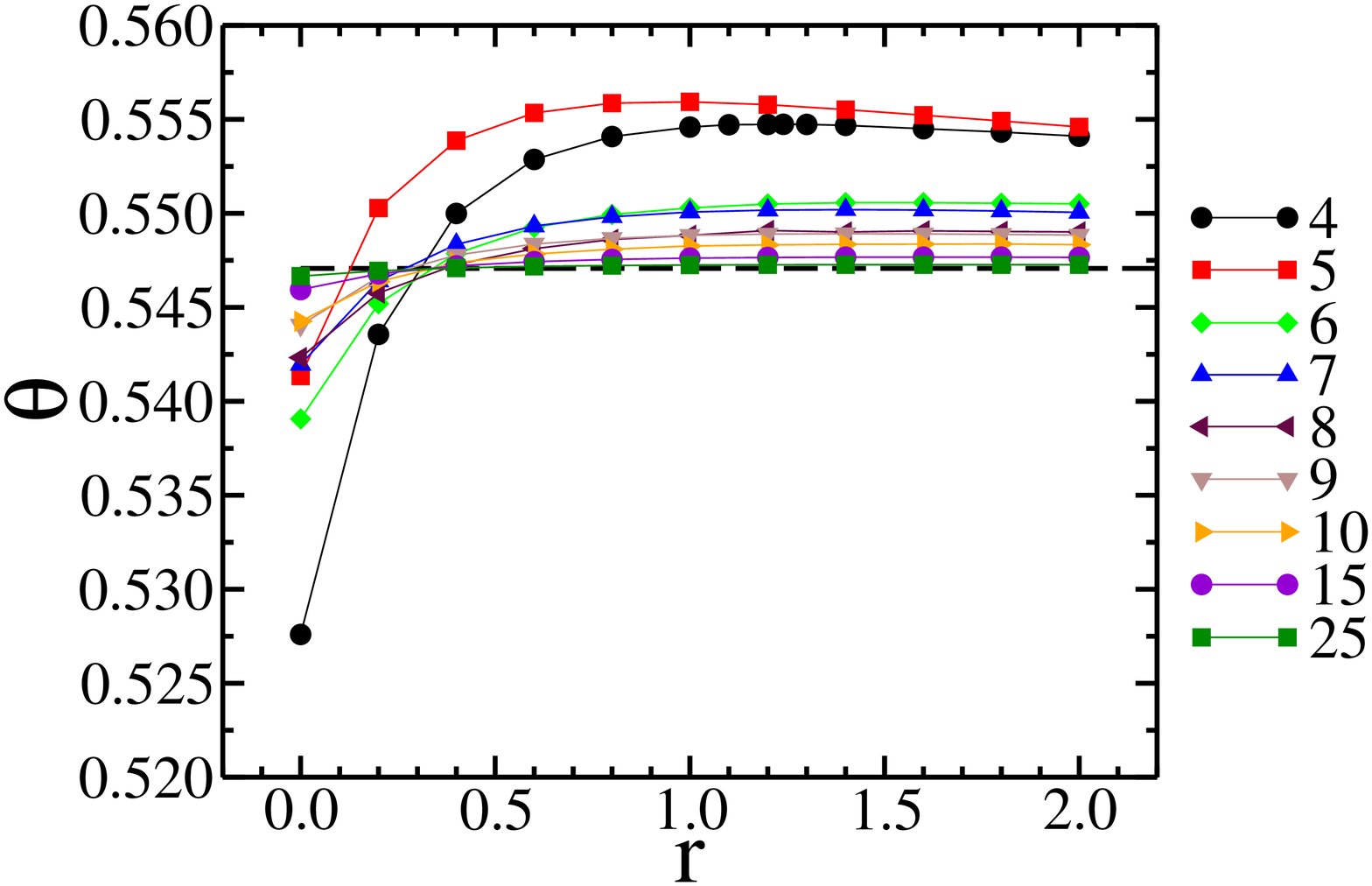}}
  }
  \caption{The dependence of the packing fraction on the radius of rounding for the equilateral triangle (a) and other regular polygons (b). Dots correspond to the mean packing fractions calculated from generated random packings and lines guide the eye. The dashed line in panel (b) corresponds to the packing fraction of disks $\theta = 0.547067$.}
  \label{fig:theta}
\end{figure}

From previous studies \cite{Zhang2018, Ciesla2014}, the packing fraction for regular polygons is the smallest for equilateral triangles: $\theta_3 = 0.525892 \pm 0.000064$ \cite{Ciesla2019}, and grows with the increasing number of polygon sides, approaching the packing density of disks $\theta_d = 0.547067 \pm 0.000003$ \cite{Ciesla2018}. Interestingly, for rounded polygons, an opposite behaviour is observed -- see Fig.~\ref{fig:theta}.
Even for relatively small rounding, the packing fractions observed are higher than $\theta_d$. For example, in case of rounded equilateral triangles the packing fraction of disks was reached around $r=0.075$. For pentagons and $r=0.2$, we observed $\theta_5(0.2) = 0.550279 \pm 0.000016$. For other shapes studied, density $\theta_d$ was exceeded for $r \ge 0.4$. It is surprising, as according to the image in Fig.~\ref{fig:5710gons}, rounded decagons for $r=1$ are visually almost indistinguishable from disks. Here, even for rounded 25-gons with $r=2$, the packing fraction is $0.547269 \pm 0.000016$, which exceeds the packing fraction of disks by a value over ten times larger than the statistical error. It shows that the local minimum of packing fraction in a shape space predicted for the most symmetric shape \cite{Baule2013, Baule2014} is really narrow. Moreover, for moderate to high rounding the highest packing fraction is observed for equilateral triangles and in general it decreases with the increasing number of polygon sides. 

Having fitted quadratic function around the maximum we found that for rounded triangles the highest packing fraction is expected near $r_3 = 0.68$ and equals $\theta_{3, \mathrm{max}} \approx 0.577$, while for rounded squares and pentagons $r_4 = 1.24$, $\theta_{4,\mathrm{max}} \approx 0.555$, and $r_5 = 0.98$, $\theta_{5,\mathrm{max}} \approx 0.556$, respectively. For other polygons, the maximum is too wide to precisely estimate its position.

To summarise, for moderate to large rounding, the densities of packings built of rounded regular polygons are higher than the density of disks' packing. On the other hand, the densities of packings built of regular polygons are smaller than this value. Rounding helps to achieve denser packings because it lowers the average excluded area blocked by a single shape. This area is measured by the parameter $B_2$ of the virial expansion \cite{Ricci1992}. For convex anisotropic figures, it has a simple form
\begin{equation}
    B_2 = 1 + \frac{P^2}{4\pi S_P},
\end{equation}
where $P$ is the perimeter and $S_P$ the surface area of the figure. One can check that the rounding growth lowers $B_2$. The value of $B_2$ is significant at the beginning of packing generation when shapes are typically far away from each other. The shape with the lowest value of $B_2=2$ is the disk. On the other hand, elongated shapes may form denser packing than disks if they are aligned in parallel. RSA favors such an alignment close to saturation limit, where neighbours restrict possible orientations of subsequent objects. Competition between these two effects makes the densest RSA packing is observed for slightly elongated figures, with relatively small $B_2$ and parallel alignment \cite{Haiduk2018, Vigil1989, Ciesla2016}. For this reason, the densest packing among all studied here is built of rounded equilateral triangles. It is worth to note here that a similar reasoning applies to other kinds of random packings. For example for three-dimensional random close packings, the highest packing fraction is observed for moderately anisotropic ellipsoids \cite{Donev2004, Man2005} or spherocylinders \cite{Zhao2012, Ferreiro2014}.

The obtained data show another interesting effect -- contrary to the general trend, the packing fractions for rounded squares are smaller than for rounded pentagons. Moreover, for small rounding, the packing fraction for rounded polygons with an odd number of sides grows faster than for even-sided rounded polygons (see tables in \ref{sec:data}). On the other hand, for larger rounding, it also decreases slightly faster. For example, for $r=2$, the packing fractions are equal to $0.9868$, $0.9988$ and $0.9976$ of their maximal values for rounded triangles, squares and pentagons, respectively. For convenience, all packing fractions measured are collected in tables in \ref{sec:data}.

\begin{figure}
    \centering{
    \subfigure[]{\includegraphics[width=0.45\linewidth]{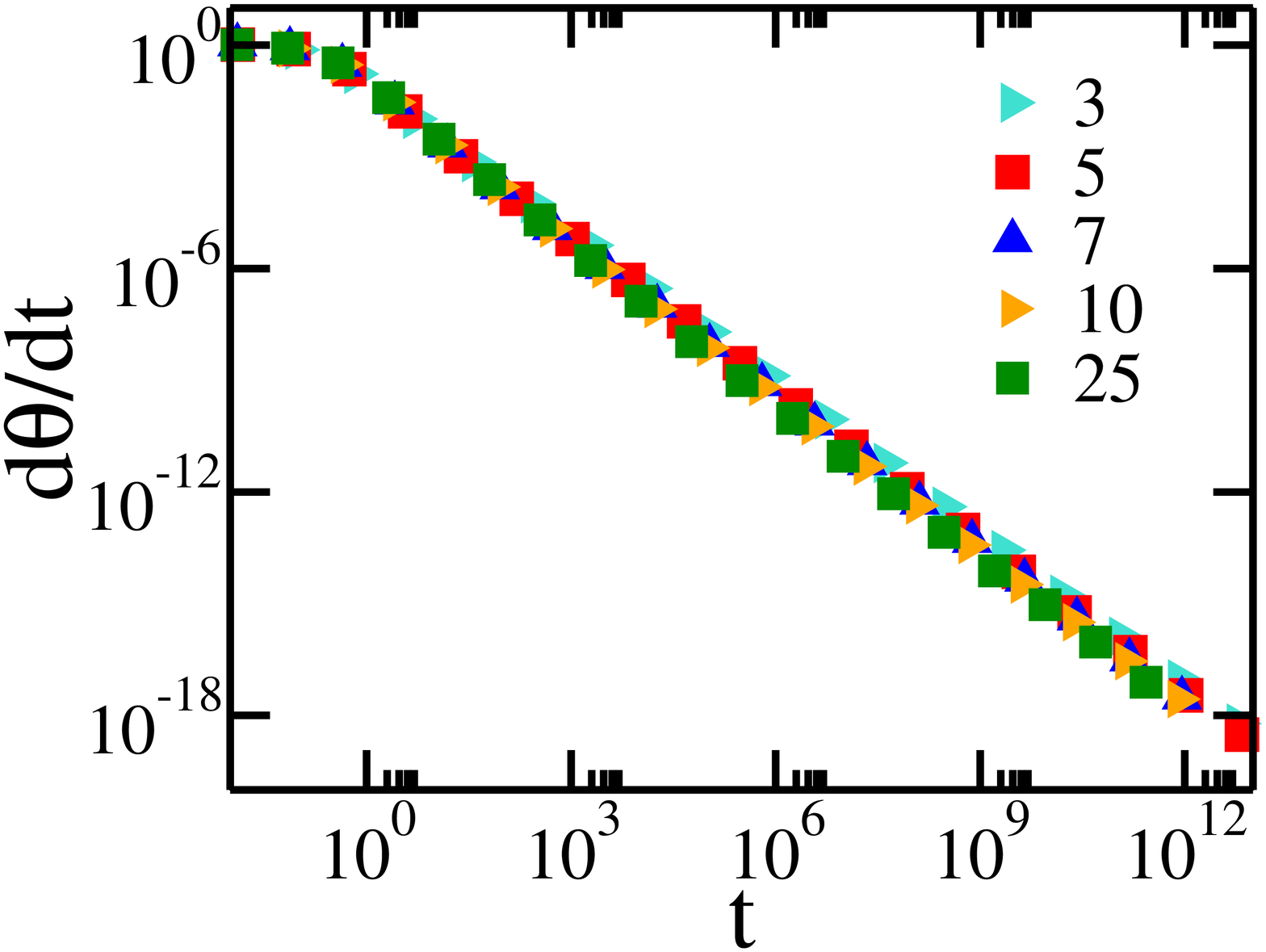}}
    \subfigure[]{\includegraphics[width=0.53\linewidth]{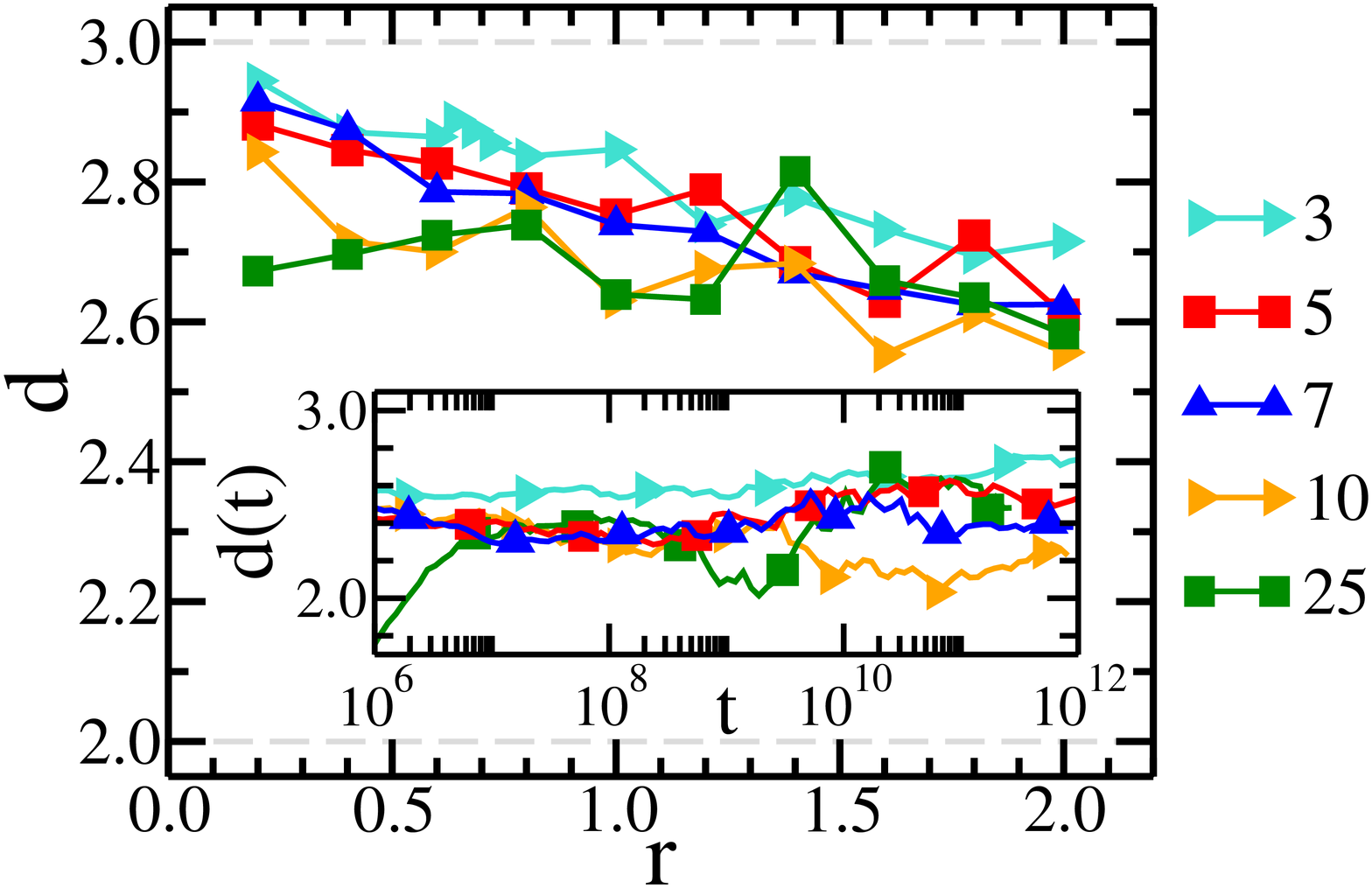}}
	}
    \caption{(a) The dependence of $\text{d}\theta/\text{d}t$ on $t$ for rounding radii giving the highest packing fractions. (b) The dependence of the power-law exponent $d$ on $r$ for a selection of studied shapes. Grey dashed lines highlight $d = 2$ and $d = 3$. Moreover, the inset shows how $d$ varies when performing $[t/100, t]$ fits with different $t$ for the same rounding radii as in (a).}
    \label{fig:d_exponent}
\end{figure}
For increasing roundness, all regular polygons approach the shape of a disk. Therefore, it is worth checking whether it is reflected in the kinetics of packing growth. For most isotropic and anisotropic shapes \cite{Zhang2013, Haiduk2018, Kasperek2018, Vigil1989, Talbot1989, Ciesla2018Random, Kubala2019} and for large enough number of iterations the difference between saturated and instantaneous packing fractions obeys the power-law
\begin{equation} \label{eq:feder}
    \theta - \theta(t) = A t^{-\frac{1}{d}},
\end{equation}
known as the Feder law \cite{Feder1980}. There $\theta \equiv \theta(\infty)$, $t = N S_P/S$ is the so-called dimensionless time, invariant under scaling of the system, $N$ is the number of iterations, $S_P$ is the area of the particle and $S$ is the area of the box. Parameter $d$ is traditionally identified with the number of particle's degrees of freedom \cite{Hinrichsen1986}, which agrees with the results for most two-dimensional shapes \cite{Zhang2013, Haiduk2018, Kasperek2018}, but seems to be violated for most three-dimensional ones \cite{Ciesla2018Random, Kubala2019}.

The parameter $d$ for the shapes studied is shown in Fig.~\ref{fig:d_exponent}b. It is determined using a linear fit to $[t/100,t]$ for large $t$ on a $\text{d}\theta/\text{d}t (t)$ log-log plot (Fig.~\ref{fig:d_exponent}a). The values of $d$ are within the $(2.5,\, 3.0)$ range and, in general, they decrease slightly with the growth of the number of sides or $r$. Interestingly, even for 25-gons with maximal $r$, the deviation from $d = 2$ is prominent, even though they are practically indistinguishable from disks (Fig. ~\ref{fig:5710gons}). This high sensitivity of $d$ to even a minute anisotropy was also observed for spherocylinders, ellipses \cite{Ciesla2016} and dimers \cite{Ciesla2014dimers}.

\begin{figure}
	\centering{
	\subfigure[]{\includegraphics[width=0.49\linewidth]{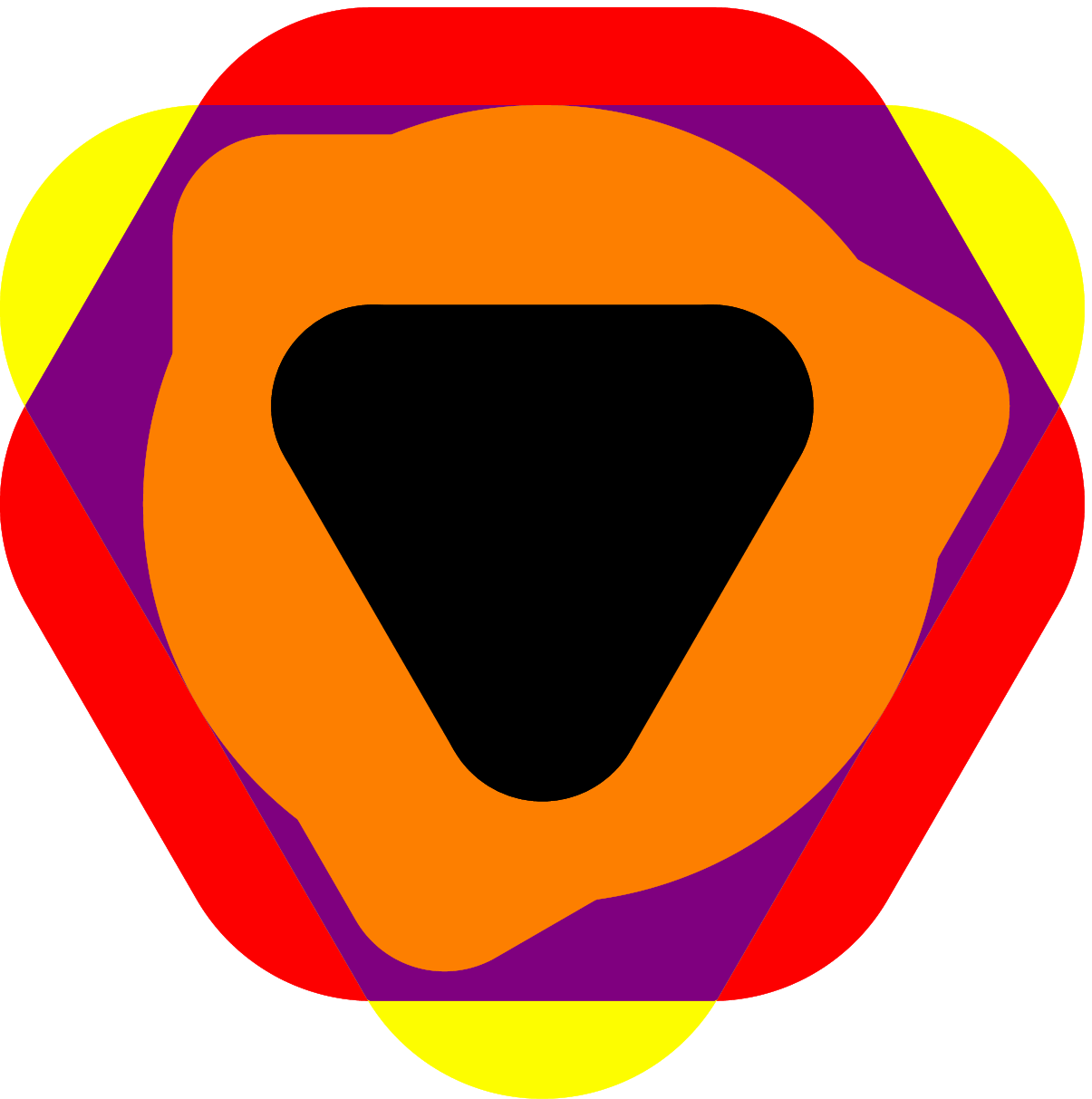}}
	\subfigure[]{\includegraphics[width=0.49\linewidth]{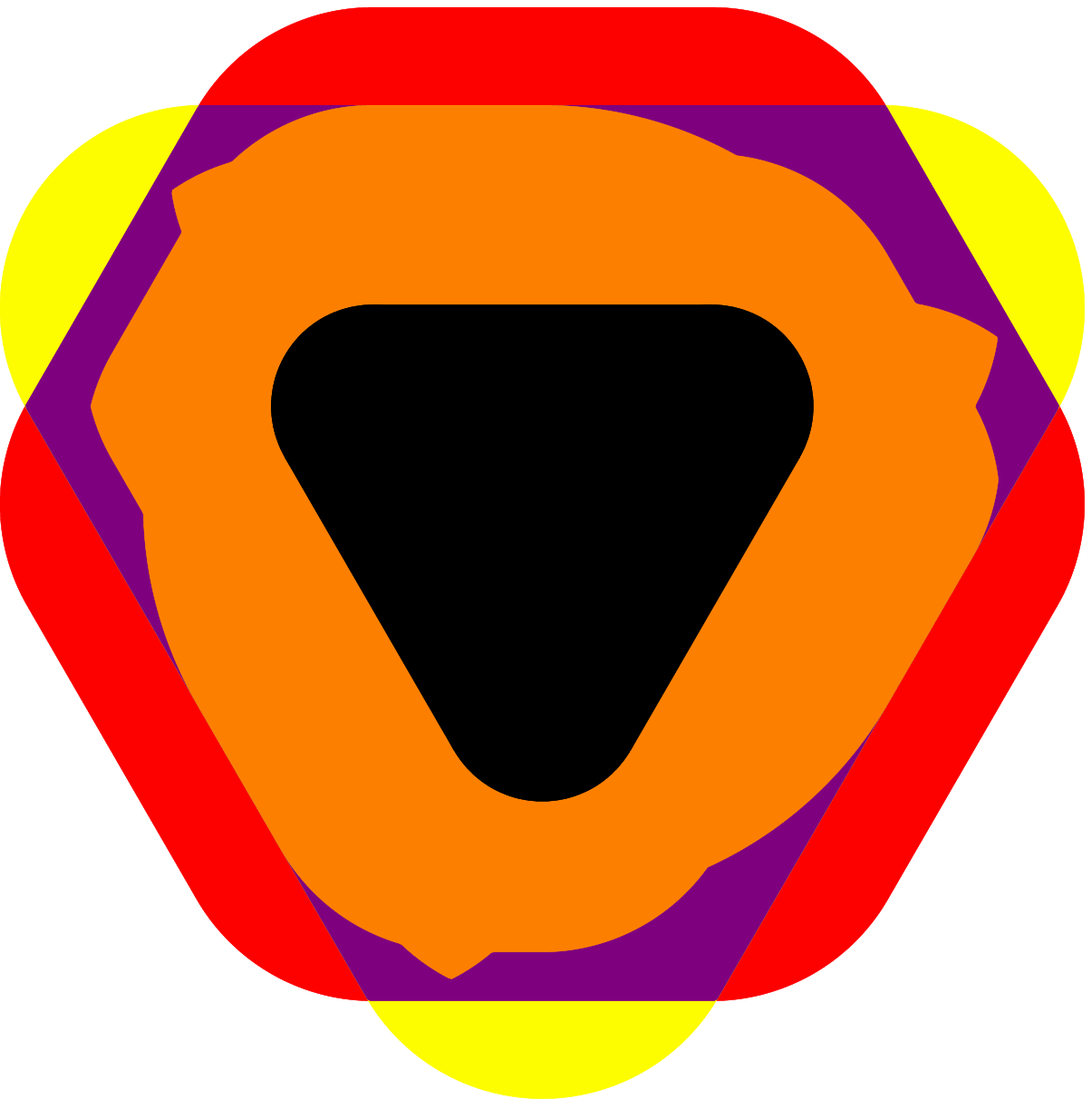}}
	}
	\caption{Exclusion zones in the algebraic approach (a) and the geometric approach (b) for the rounded equilateral triangle (colour online). The black area is the particle. The red and yellow areas correspond to exclusion zones for second particle's orientations $\alpha = 0$ and $\alpha = \pi/3$, respectively. The orange area estimates the exclusion zone for the whole range $[0,\pi/3]$. Note that the exclusion zone (b) is not symmetric. It is due to approximating voxel arcs by rectangular boundings whose shape depends not only on the central angle but also on both endpoints.}
    \label{fig:ex_zone_comparison}
\end{figure}
\subsection{Algorithm efficiency}
\begin{figure}
    \centering{
    \includegraphics[width=0.7\linewidth]{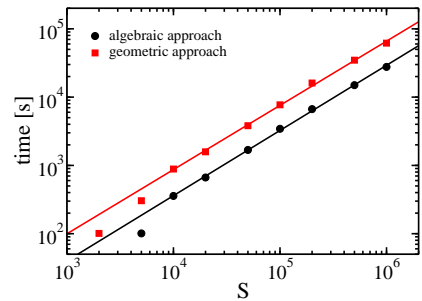}
    }
    \caption{The dependence of the time needed to generate $100$ different packings built of rounded pentagons on the packing size for the algebraic and geometric approaches. Dots are the data, and solid lines are power fits $t = 0.055 S^{0.95}$ and $0.15 S^{0.94}$ for geometric and algebraic approaches, respectively.}
    \label{fig:time}
\end{figure}

As there were two independent voxel elimination schemes provided, it is interesting to determine which one is more efficient. The effectiveness of voxel elimination depends, among other factors, on how well the maximal exclusion zones (\ref{eq:max_exzone}) are approximated and, as a result, how many voxels are eliminated during the early stage. Although the algebraic approach does not explicitly utilise the excluded volume notion, it emerges due to the intersection criterion examined for the voxel. The exclusion zones given by the two algorithms, for the same particle and voxel angle ranges, with voxel size tending to 0, are shown in Fig.~\ref{fig:ex_zone_comparison}. In this case, due to very rough estimations, the algebraic approach gives an exclusion zone of a smaller area than the second algorithm. Nevertheless, the former leads to over a three times faster packing generation despite a higher number of voxels before each division (Fig.~\ref{fig:time}).
On the other hand, the geometric approach seems to be much more versatile, as it can be applied not only to objects built of spherocylinders but also to other shapes built of any figures, for which we know the recipe to construct the exclusion zone. 

\begin{figure}
    \centering{
    \includegraphics[width=0.7\linewidth]{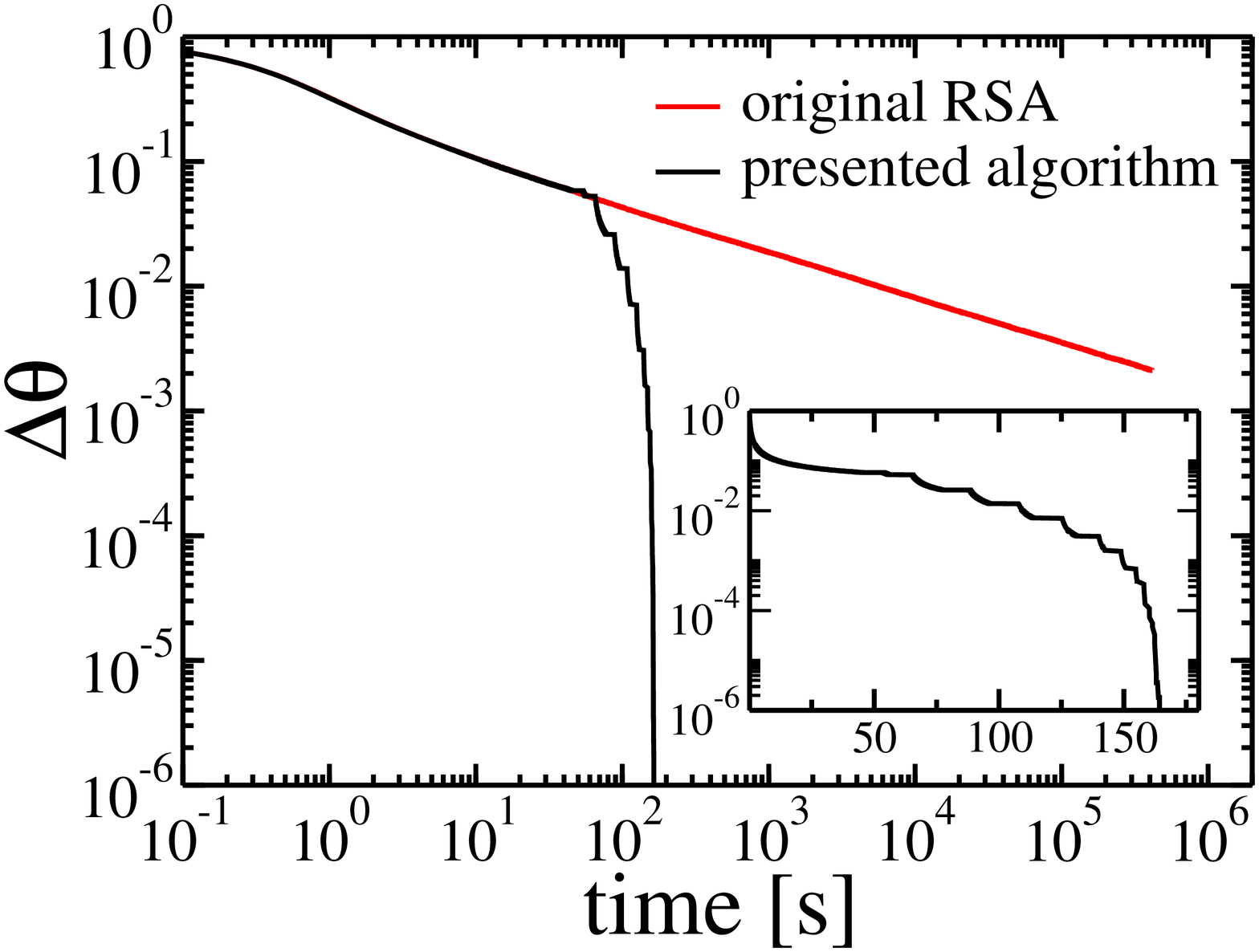}
    }
    \caption{The dependence of the relative distance to saturation for a large packing ($S=10^6$) built of rounded pentagons on time for classical RSA and the algorithm presented here using the arithmetic approach to eliminate voxels. The inset shows the same data for the presented algorithm, however in a linear time scale.}
    \label{fig:efficiency}
\end{figure}

Finally, we compare the algorithm developed and the classical version of RSA (see Fig.~\ref{fig:efficiency}).
The plot presents the dependence of $\Delta \theta = (\theta - \theta(t))/\theta$ on the simulation time $t$. Our algorithm generates one strictly saturated packing in $164$ seconds, whereas the packing generated by the classical RSA after almost five days of computations still had regions large enough to place over one thousand additional figures. The reason for such a high inefficiency is that for a nearly saturated packing the regions in $(x, y, \alpha)$ configuration space not covered by exclusion zones are so small, that they may require huge number of iteration to be sampled. In the presented case this number is of the order of $10^{20}$ and it grows exponentially with the packing size \cite{Ciesla2017}. As the shape of exclusion zones, especially in the angular direction, is very intricate, standard methods are insufficient \cite{Torquato2006}. Note the steps in the black line. They correspond to voxel division, analysis and elimination phase of the algorithm. In this particular case, there were $21$ such phases, and they took $79$ seconds, which is almost half of the total packing generation time. The packing size was $10^6$, and the initial spatial size of each voxel was $0.5$, while the initial angular size was $\pi /10$. Voxels were split after approximately $4 \times 10^4$ consecutive unsuccessful tries of adding a figure to the packing. All the above tests were run on a computer equipped with 4-core Intel i7 920 CPU running at 2.67~GHz and 12~GB of RAM.
\section{Summary}
We described a two-variant algorithm for generating strictly saturated RSA packings built of rounded polygons. It bases on tracking the space where adding the next particle is possible. The algorithm was used to determine the densities of saturated packings built of regular rounded polygons. The highest packing fraction was obtained for the rounded equilateral triangle of the rounding radius equal approximately $0.7$ of the side length. Its value is $0.577248 \pm 0.000017$. With the increasing number of polygon sides, the packing fraction approaches the density of a packing built of disks $\theta_d = 0.547067 \pm 0.000003$, but even for a packing built of 25-gons, the density $\theta_{25}(r=2) = 0.547269 \pm 0.000016$ is slightly higher than that. It means that even unnoticeable changes of figure shape may follow different RSA packing fractions.

The method presented can be used to study the theoretical properties of random packing and effectively model monolayers arising in irreversible adsorption processes of a large variety of different adsorbates.
\section*{Acknowledgments}
This work was supported by grant no.\ 2016/23/B/ST3/01145 of the National Science Centre, Poland. Numerical simulations were carried out with the support of the Interdisciplinary Centre for Mathematical and Computational Modeling (ICM) at the University of Warsaw under grant no.\ GB76-1.
\appendix
\section{Numerical data}
\label{sec:data}
Below are the data obtained from numerical simulations.
\begin{table*}[ht]
  \setlength{\tabcolsep}{5pt}
  \renewcommand{\arraystretch}{1}
  \begin{tabular}{r|r|r|r|r|r|r|}
  $r$    & $\theta$ ($n = 3$) & $\theta$ ($n = 4$) & $\theta$ ($n = 5$) & $\theta$ ($n = 6$) & $\theta$ ($n = 7$)  \\
  \hline
  0.00 & 0.525820(66)$^*$  & 0.527594(70)$^*$  & 0.541344(72)$^*$  & 0.539060(95)$^*$  & 0.541959(124)$^*$  \\
  0.05 & 0.541286(21) & & & & \\
  0.07 & 0.546035(20) & & & & \\
  0.08 & 0.548162(20) & & & & \\
  0.10 & 0.551992(20) & & & & \\
  0.15 & 0.559522(17) & & & & \\
  0.20 & 0.564874(17)   & 0.543589(19)   & 0.550279(16)   & 0.545123(18)   & 0.546400(18)    \\
  0.40 & 0.574829(17)   & 0.549994(20)   & 0.553867(16)   & 0.547837(17)   & 0.548346(17)    \\
  0.60 & 0.577150(17)   & 0.552872(19)   & 0.555341(16)   & 0.549193(17)   & 0.549326(17)    \\
  0.64 & 0.577225(16)   &                &                &                &                 \\
  0.68 & 0.577248(17)   &                &                &                &                 \\
  0.72 & 0.577221(17)   &                &                &                &                 \\
  0.80 & 0.577007(18)   & 0.554087(19)   & 0.555864(17)   & 0.549923(16)   & 0.549819(16)    \\
  1.00 & 0.576024(17)   & 0.554606(19)   & 0.555934(16)   & 0.550291(18)   & 0.550067(17)    \\
  1.10 &                & 0.554715(15)   &                &                &                 \\
  1.20 & 0.574733(16)   & 0.554740(18)   & 0.555780(16)   & 0.550487(17)   & 0.550169(17)    \\
  1.24 &                & 0.554735(17)   &                &                &                 \\
  1.30 &                & 0.554732(16)   &                &                &                 \\
  1.40 & 0.573389(16)   & 0.554683(17)   & 0.555522(17)   & 0.550558(17)   & 0.550195(17)    \\
  1.60 & 0.572054(17)   & 0.554501(17)   & 0.555222(17)   & 0.550567(18)   & 0.550175(16)    \\
  1.80 & 0.570783(17)   & 0.554299(18)   & 0.554912(17)   & 0.550534(17)   & 0.550125(16)    \\
  2.00 & 0.569605(16)   & 0.554103(17)   & 0.554598(17)   & 0.550449(16)   & 0.550051(16)    \\
  \hline
  \end{tabular}
  \caption{Mean saturated packing fractions for rounded regular polygons. Numbers in angular brackets correspond to standard deviations of the mean value at two last digits. Values marked with asterisk were taken from \cite{Zhang2018}.}
\end{table*}
\begin{table*}[ht]
  \setlength{\tabcolsep}{5pt}
  \renewcommand{\arraystretch}{1}
  \begin{tabular}{r|r|r|r|r|r|}
  $r$   & $\theta$ ($n = 8$) & $\theta$ ($n = 9$) & $\theta$ ($n = 10$) & $\theta$ ($n = 15$) & $\theta$ ($n = 25$) \\
  \hline
  0.0 & 0.542328(98)$^*$  & 0.544059(89)$^*$  & 0.54426(12)$^*$    & 0.545942(17)   & 0.546655(16) \\
  0.2 & 0.545730(19)   & 0.546609(17)   & 0.546347(17)    & 0.546784(17)   & 0.546941(17) \\
  0.4 & 0.547279(19)   & 0.547775(18)   & 0.547308(18)    & 0.547204(17)   & 0.547094(17) \\
  0.6 & 0.548102(17)   & 0.548366(18)   & 0.547814(17)    & 0.547430(17)   & 0.547181(17) \\
  0.8 & 0.548560(17)   & 0.548670(18)   & 0.548089(16)    & 0.547551(17)   & 0.547224(17) \\
  1.0 & 0.548803(17)   & 0.548828(18)   & 0.548267(18)    & 0.547618(17)   & 0.547250(17) \\
  1.2 & 0.548930(19)   & 0.548896(18)   & 0.548320(16)    & 0.547651(17)   & 0.547265(17) \\
  1.4 & 0.549004(17)   & 0.548914(17)   & 0.548353(16)    & 0.547671(17)   & 0.547270(17) \\
  1.6 & 0.549020(17)   & 0.548903(17)   & 0.548359(16)    & 0.547671(17)   & 0.547269(17) \\
  1.8 & 0.549034(19)   & 0.548882(18)   & 0.548329(16)    & 0.547667(17)   & 0.547269(17) \\
  2.0 & 0.549011(19)   & 0.548843(17)   & 0.548329(16)    & 0.547659(17)   & 0.547269(16) \\
  \hline
  \end{tabular}
  \caption{Mean saturated packing fractions for rounded regular polygons (continuation). Numbers in angular brackets correspond to standard deviations of the mean value at two last digits. Values marked with asterisk were taken from \cite{Zhang2018}.}
\end{table*}

The source code of the implementation of described algorithm is available at \linebreak \url{https://github.com/misiekc/polygonRSA}.
\bibliographystyle{apsrev4-1}
\bibliography{main}
\end{document}